\documentclass[reprint,
superscriptaddress,
amsmath,
amssymb,
longbibliography,
floatfix,
aps,
pra,
footinbib]{revtex4-2}

\usepackage{graphicx}
\usepackage{dcolumn} 
\usepackage{xcolor}
\usepackage[absolute]{textpos}
\usepackage{bm}
\usepackage{braket}
\definecolor{darkblue}{rgb}{0,0.3,0.7}
\usepackage{algorithm, dsfont}
\usepackage{algpseudocode}
\usepackage{mathdots}
\usepackage{upgreek}
\usepackage{verbatim}
\usepackage{lipsum}
\usepackage{mathtools}
\usepackage{bibunits}
\defaultbibliographystyle{apsrev4-2}
\usepackage{hyperref, comment}
\hypersetup{
colorlinks=true,
linkcolor=darkblue,
filecolor=blue,
citecolor=darkblue,  
urlcolor=darkblue,
}

\definecolor{teal}{RGB}{26,157,150}

\definecolor{ornlblue}{RGB}{80,145,205}

\usepackage{tikz}
\usetikzlibrary{positioning, arrows.meta, shapes, calc, decorations.pathmorphing}
\usepackage[normalem]{ulem} 

\usepackage[T1]{fontenc}     
\usepackage[utf8]{inputenc}  
\usepackage[scaled]{sourcecodepro} 
\newcommand{\cF}{\mathcal{F}}
\newcommand{\cP}{\mathcal{P}}

\newcommand{\cE}{\mathcal{E}}
\newcommand{\hcE}{\hat{\mathcal{E}}}
\newcommand{\hcF}{\hat{\mathcal{F}}}
\newcommand{\hcP}{\hat{\mathcal{P}}}
\newcommand{\hrho}{\hat{\rho}}
\newcommand{\hPhi}{\hat{\Phi}}
\newcommand{\bN}{\mathbf{N}}
\newcommand{\bx}{\mathbf{x}}
\def\bcD{{\bm{\mathcal{D}}}}
\DeclareMathOperator{\Tr}{Tr}
\usepackage[capitalize]{cleveref}
\crefname{section}{Sec.}{Secs.}
\Crefname{section}{Section}{Sections}

\crefrangelabelformat{section}{#3#1#4--#5\crefstripprefix{#1}{#2}#6}
\crefrangelabelformat{figure}{#3#1#4--#5\crefstripprefix{#1}{#2}#6}
\crefrangelabelformat{equation}{(#3#1#4--#5\crefstripprefix{#1}{#2}#6)}

\crefmultiformat{equation}%
{\edef\crefstripprefixinfo{#1}Eqs.~(#2#1#3}%
{,#2\crefstripprefix{\crefstripprefixinfo}{#1}#3)}%
{,#2\crefstripprefix{\crefstripprefixinfo}{#1}#3}%
{,#2\crefstripprefix{\crefstripprefixinfo}{#1}#3)}

\begin{document}
\title{A quantum game of telephone}
\author{Arefur Rahman}
\author{Matthew L. Stevens}
\affiliation{Elmore Family School of Electrical and Computer Engineering and Purdue Quantum Science and Engineering Institute, Purdue University, West Lafayette, Indiana 47907, USA}

\author{Cory M. Nunn}
\affiliation{National Institute of Standards and Technology, Gaithersburg, Maryland 20899, USA}
\affiliation{Joint Quantum Institute and University of Maryland, College Park, Maryland 20742, USA}

\author{Daniel~E. Jones}
\affiliation{DEVCOM US Army Research Laboratory, Adelphi, Maryland 20783, USA}
\affiliation{National Institute of Standards and Technology, Gaithersburg, Maryland 20899, USA}

\author{Brian T. Kirby}
\affiliation{DEVCOM US Army Research Laboratory, Adelphi, Maryland 20783, USA}
\affiliation{Tulane University, New Orleans, LA 70118, USA}

\author{Joseph~M. Lukens}
\email{jlukens@purdue.edu}
\affiliation{Elmore Family School of Electrical and Computer Engineering and Purdue Quantum Science and Engineering Institute, Purdue University, West Lafayette, Indiana 47907, USA}
\affiliation{Quantum Information Science Section, Oak Ridge National Laboratory, Oak Ridge, Tennessee 37831, USA}

\date{\today}

\begin{abstract}
Characterizing multinode quantum networks without ubiquitous local entanglement sources presents significant experimental challenges. 
We introduce ``quantum telephone,'' an iterative ancilla-assisted process tomography protocol that leverages intermediate detection events, effectively treating previously probed links as ancillae for subsequent links.
Implemented on a deployed multinode fiber network, we observe noisy intermediate channels create fundamental parameter degeneracies that compound inference errors under sequential estimation.
Counterintuitively, the inclusion of downstream near-unitary channels provides boundary constraints that, when used in tandem with global inference, can resolve ambiguity in channels earlier in the sequence.
By compensating for localized information loss, this approach obviates the strict full-rank requirements of standard ancilla-assisted process tomography, even when intermediate states become completely depolarized.
Overall, quantum telephone offers a hardware-efficient and information-maximizing path toward characterizing complex quantum networks with limited resources.
\end{abstract}

\maketitle

\section{Introduction}
Large-scale metropolitan quantum network testbeds are being constructed throughout the globe~\cite{Alshowkan2021, neumann2022continuous, chung2022design, Liu2024, Stolk2024, mckenzie2024clock,craddock2024automated,kucera2024demonstration,bersin2024development, Chapman2024}, composed of nodes that generate, manipulate, and measure quantum states, and channels (like optical fiber) over which these states are transmitted~\cite{kimble2008quantum,wehner2018quantum,azuma2023quantum, Lukens2025}.
Quantum channel characterization is essential for their operation and, ultimately, their ability to perform tasks that are unavailable to classical networks.
A significant complication in deployed channels is time-dependent drift~\cite{mckenzie2024clock, ding17polarization}, which necessitates rapid approaches for both assessing the network's operational status and implementing real-time compensation systems~\cite{Xavier2008,treiber2009fully,Xavier2009,neumann2022continuous,Peranic2023,craddock2024automated,kucera2024demonstration,Chapman2024,bersin2024development,Sena2025,Stevens2026,Shi2026}

Quantum process tomography is a principled approach for completely characterizing an unknown quantum channel, historically based on quantum state tomography (QST) of outputs to a complete set of linearly independent inputs~\cite{chuang1997prescription,poyatos1997complete,bongioanni2010experimental}.
While effective, it is often more convenient to use ancilla-assisted process tomography (AAPT) in quantum networking settings, as this can leverage entanglement sources and detectors already in place~\cite{d2001quantum,altepeter2003ancilla}.
To date, AAPT of deployed quantum links has been limited to a relatively small number of lightpaths, allowing for characterization to be performed independently and with quasi-unitary ancillae~\cite{rahman2025deployed,Stevens2026}.
However, as quantum networks scale, it is unlikely that every node will be equipped with the local entanglement sources necessary for AAPT.

Existing frameworks for quantum network tomography (QNT) tackle the lack of ubiquitous sources by deducing interior channel noise strictly from end-to-end measurements, assuming specialized channel forms that reduce the number of unknowns for tractable inference~\cite{de2022quantum,de2024quantum,wang2025quantum,zheng2026quantum,navas2026loss}.
Yet practically speaking, emerging deployed networks operate in a middle ground between true AAPT and QNT: while they cannot support high-fidelity entanglement generation at every node, they frequently possess intermediate detection and measurement capabilities.

\begin{figure}[tb!]\centering
\includegraphics[width=\columnwidth]{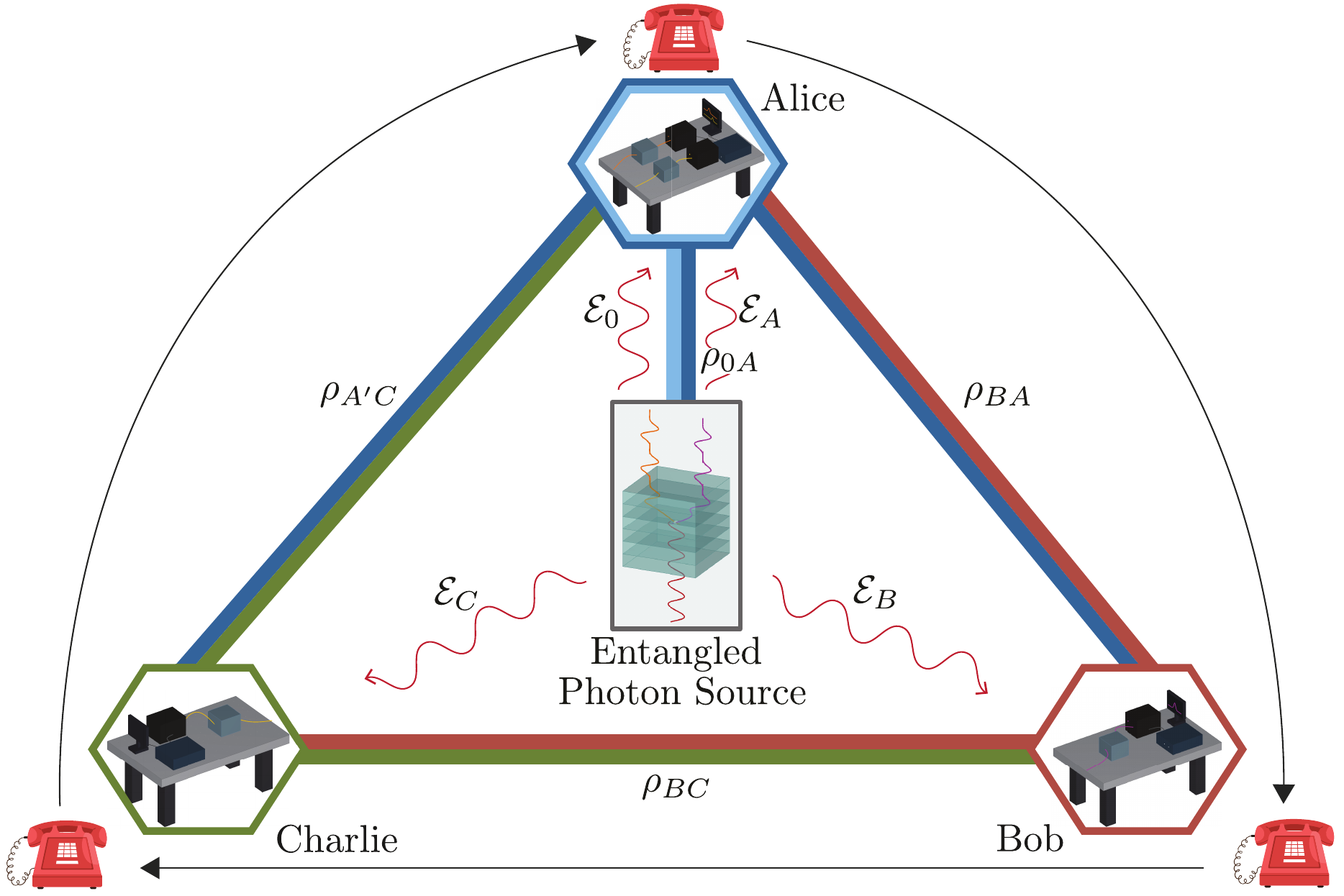}
    \caption{Schematic of a three-node quantum telephone experiment. 
    A central nonlinear source generates polarization-entangled photon pairs, which are distributed as bipartite states $\{\rho_{0A}, \rho_{BA}, \rho_{BC}, \rho_{A^{'}C}\}$ (colored lines) after optical channels $\{\mathcal{E}_0, \mathcal{E}_A, \mathcal{E}_B, \mathcal{E}_C\}$ (crimson arrows) to three nodes: Alice ($A/A'$), Bob ($B$), and Charlie ($C$).
    Bootstrapping on a single ancilla channel at Alice, all unknown quantum channels are then inferred using the measured data, the accuracy of which depends on both the ground truth channels and the estimation procedure.}
    \label{fig:concept}
\end{figure}

In this work, we address this middle ground by introducing ``quantum telephone,'' an iterative AAPT-inspired approach to network characterization where previously probed channels act as ancillae for subsequent links. 
We demonstrate this protocol on a deployed quantum network in a circular topology, benchmarking the same target channel initially and again after chaining completely around the network (\cref{fig:concept}). 
Much like the children's game of telephone, we find that sequential quantum probing is highly vulnerable to error compounding: noisy channels obscure whether observed effects originate from the ancilla, the target, or both.

However, we show how to subvert the rules of this traditional game: like an observer collecting and cross-referencing every player’s intermediate message, we utilize a global inference algorithm that evaluates all network measurements simultaneously. When combined with downstream near-unitary links, this global evaluation resolves upstream parameter degeneracies and recovers quantum channels previously masked by noise.
By mitigating the rigid compounding of cascaded errors, this framework establishes a robust methodology for characterizing complex quantum network architectures that is hardware-efficient, circumvents the need for ubiquitous entanglement sources, and can unlock hidden tomographic data that sequential methods miss.

\section{Quantum telephone}
Consider a quantum network with $T$ unknown quantum channels $\{\cE_1,...,\cE_T\}$, each acting on a single $d$-dimensional qudit with density matrix $\sigma$ as 
\begin{equation}
\label{eq1:cptp_map}
    \cE_t(\sigma) = \sum_{k=1}^R A_{tk} \sigma A^{\dagger}_{tk},
\end{equation}
where $t\in\{1,...,T\}\equiv[T]$, $R \le d^2$ defines the Choi rank, and $A_{tk}$ are Kraus operators such that $\sum_k A^{\dagger}_{tk} A_{tk} = \mathds{1}$.
Using the standard Choi-Jamiołkowski isomorphism~\cite{Gilchrist2005,Wilde2017}, we can map $\cE_t$ to its Choi state $\Phi_t$ and use both interchangeably to refer to ``the'' channel $t$.

At a single source node, two qudit subsystems are prepared in the joint entangled state $\rho$ and initially characterized by estimator $\hrho$, where we reserve carets for inferred quantities to distinguish them from the ground truth $\sigma$ ($\rho$) for a single-qudit (two-qudit) state.
Quantum telephone probes $T$ unknown channels in overlapping pairs, leading to the $T$ output states
\begin{equation}
\label{eq:recursive}
\rho_{t-1,t} = (\cE_{t-1}\otimes\cE_t)(\rho),
\end{equation}
with $\cE_0=\mathds{1}$. Measuring each state $\rho_{t-1,t}$ in a tomographically complete set of $S$ two-qudit projections $\{\ket{\varphi_s}\}_{s\in [S]}$ yields a vector of coincidence counts $\bN_t=(N_{t1},...,N_{tS})$.

Leveraging the theory behind AAPT, $\bN_1$ therefore provides complete information on $\cE_1$.
The intuition behind quantum telephone is to bootstrap this knowledge to recover subsequent channels not probed via AAPT. Conceptually, the logic flows sequentially
\begin{equation}
\label{eq:sequential}
\underbrace{\{\bN_1\} \rightarrow \hcE_1, \; \{\hcE_1, \bN_2\} \rightarrow \hcE_2, \; \cdots, \; \{\hcE_{T-1}, \bN_T\} \rightarrow \hcE_T}_{\text{sequential quantum telephone}},
\end{equation}
where the arrows stand for inference in which the preceding information is combined to yield an estimate for the next channel.

Because noisy channels can remove correlations from the input $\rho$, only the estimation of $\hcE_1$ is a standard instantiation of AAPT, which requires the bipartite probe state to possess full operator Schmidt rank for unique identification of an arbitrary quantum channel~\cite{d2001quantum, altepeter2003ancilla, lie2023faithfulness}; entanglement is sufficient but not necessary to meet this condition, sometimes called faithfulness.
Consequently, sequential quantum telephone requires all intermediate channels to preserve this full-rank condition, which is not guaranteed in general.

However, as we will show experimentally, global inference methods can overcome localized failures of this requirement by retroactively applying tomographic constraints from highly correlated downstream measurements, via an inference procedure schematized  as
\begin{equation}
\label{eq:global}
\underbrace{\{\bN_1,...,\bN_T\}\rightarrow\{\hcE_1,...,\hcE_T\}}_\text{global quantum telephone}.
\end{equation}
Because all $T$ channels are estimated collectively, the number of unknown parameters per single inference grows like $\mathcal{O}(Td^4)$---compared to $\mathcal{O}(d^4)$ in \cref{eq:sequential}---which can significantly increase the total computational load, particularly for time-intensive Bayesian techniques~\cite{Blume-Kohout2010, Lukens2020b, Nguyen2025}.
However, the explicit enforcement of cross-measurement consistency in \cref{eq:global} unlocks the opportunity for higher accuracy.

\section{Experimental implementation}
We test quantum telephone with polarization qubits on a three-node quantum network of spatially separated rooms---Alice $A$, Bob $B$, and Charlie $C$---in the subbasement of the Material Science and Electrical Engineering (MSEE) building at Purdue University~\cite{Stevens2026}.
Though not required for quantum telephone \emph{per se}, we summon the spirit of the traditional game of telephone and infer the initial channel twice, yielding an end-to-end benchmark of the procedure's accuracy. 
In the first experiment, Charlie in \cref{fig:concept} is bypassed, and the following output states are measured:
\begin{equation}
\label{eq5:qt_maps_2n}
\begin{split}
    \rho_{0A} &= (\cE_0 \otimes \mathcal{E}_A)(\rho) \\
    \rho_{BA} &= (\mathcal{E}_B \otimes \mathcal{E}_A)(\rho) \\
    \rho_{BA'} &= (\mathcal{E}_B \otimes \mathcal{E}_{A'})(\rho).
\end{split}
\end{equation}
The second experiment considers the full three-node loop in \cref{fig:concept} and obtains data for $\rho_{0A}$, $\rho_{BA}$, and
\begin{equation}\label{eq6:qt_maps_3n}
\begin{split}
    \rho_{BC} &= (\mathcal{E}_B \otimes \mathcal{E}_C)(\rho) \\
    \rho_{A'C} &= (\mathcal{E}_{A'} \otimes \mathcal{E}_C)(\rho).
\end{split}
\end{equation}
In contrast to the truly recursive ordering of \cref{eq:recursive}, we alternate the ``ancilla'' subsystem of the bipartite state at each successive estimation; i.e., with each step the most recently probed channel is still applied to the same photon.
This procedure is adopted for ease of implementation, but otherwise makes no meaningful difference for the protocol.
Additionally, although $\cE_A$ and $\cE_{A'}$ denote the same physical channel, they are treated as entirely independent entities in the inference process.

After obtaining experimental coincidence data--- ${\bcD=\{\bN_{0A},\bN_{BA},\bN_{BA'}\}}$ for the two-node case, and ${\bcD=\{\bN_{0A},\bN_{BA},\bN_{BC},\bN_{A'C}\}}$ for three---we estimate the unknown quantum channels according to either a sequential [\cref{eq:sequential}] or global [\cref{eq:global}] procedure. 
Bayesian inference~\cite{Blume-Kohout2010} is selected due to its natural uncertainty quantification, optimality in mean squared error, and avoidance of unjustified low-rank estimators.
To sample from the high-dimensional probability distributions involved, we enlist  Markov chain Monte Carlo (MCMC) techniques: specifically, the workflow proposed in Refs.~\cite{Lukens2020b,Lu2022b}, parallelized in Ref.~\cite{Nguyen2025}, and applied to quantum process tomography in Refs.~\cite{Chapman2023,rahman2025deployed, Stevens2026}.

The general procedure follows that of Ref.~\cite{Stevens2026} except that the probability $\wp_{ts}$ of obtaining the result $\ket{\varphi_s}$ for output state $\rho_{t-1,t}$ now depends on two \emph{a priori} unknown quantum processes:
\begin{equation}
\label{eq:seqProb}
\wp_{ts}(\bx)=\braket{\varphi_s|\left[\cE_{t-1}(\bx)\otimes\cE_t(\bx)\right](\hrho)|\varphi_s},
\end{equation}
where $\hrho$ denotes the Bayesian mean of the initially estimated quantum resource state. In sequential inference, $\cE_{t-1}(\bx)$ is replaced by its previously estimated Bayesian mean $\hcE_{t-1}$, but otherwise the procedure is identical to single-channel Bayesian AAPT.
For global inference, both processes are modeled by  Kraus operators $A_{tk}(\bx)$ that depend on the unknown parameters $\bx$, and the total likelihood is the product of the $T$ likelihoods for each of the individual telephone datasets.
The procedure outputs $N$ samples $\bx^{1:N}=\{\bx^i\}_{i\in[N]}$ drawn from the posterior distribution, each parameter vector $\bx^i$ defining a quantum process $\Phi_t(\bx^i)$---or all $T$ quantum processes in the global case.
A total number of samples $N=2^{10}$ are saved from $2^{22}$ iterations with a thinning interval of $2^{12}$.
Global inference consistently requires more computation time than sequential inference: for two nodes, $50$ min versus $40$ min on a standard laptop computer; for three nodes, $70$ min versus $48$ min.

The following experiments focus on three main goals: (i) demonstrating the feasibility of quantum telephone as a whole, (ii) comparing the relative merits of sequential versus global estimation, and (iii) testing the limits of the method for noisy upstream channels.
In order to avoid complications to the analysis from residual unitary rotations, we quantify the estimated channel according to the purity $\cP=\Tr\Phi^2$, whose Bayesian mean can be computed from the MCMC samples $\bx^{1:N}$ as
\begin{equation}
\label{eq:purity}
\hcP_t =\frac{1}{N}\sum_{i=1}^N \Tr\Phi_t^2(\bx^i).
\end{equation}
Because local unitary rotations are easily and routinely compensated in deployed networks (and irrelevant to entanglement quantification), our focus on the gauge-invariant purity isolates important features of the channel quality that may otherwise be obscured by gauge-dependent metrics like fidelity. Appendix~\ref{sec:fidelityAnalysis} provides further discussion of the nuances associated with fidelity-based analysis, with comparisons to simulations results in Appendix~\ref{sec:simulations}.

As previously tested \cite{Stevens2026}, the Bob and Charlie links are highly unitary, so we take
\begin{equation}
\label{eq:BCideal}
\cE_{B(C)}^\star(\sigma)=U\sigma U^\dagger
\end{equation}
as the ideal for comparison, with purity $\cP_{B(C)}^\star=1$, where we have adopted the notation that any quantity with a hat denotes the Bayesian mean (e.g., $\hPhi_t$)  and any quantity with a superscript star denotes the ideal target (e.g., $\Phi_t^\star$).
Alice is designed to impart controllable noise, which we realize through polarization scrambling with depolarizing probability $p$:
\begin{equation}
\label{eq:Aideal}
\cE_{A(A')}^\star(\sigma)=(1-p)U\sigma U^\dagger+ \frac{p}{2}I_2,
\end{equation}
where $I_d$ denotes the $d\times d$ identity matrix. The Choi matrix of this channel has purity that reduces quadratically from unity at $p=0$ to one-fourth at $p=1$: $\cP_{A(A')}^\star=1-\frac{3p}{2}+\frac{3p^2}{4}$.
The degree to which the estimated channel purity $\hcP_t$ matches the relevant theoretical expectation $\cP_t^\star$ will form our primary metric for evaluating quantum telephone-based inference.

\section{Results}
\begin{figure}[!t]\centering
\includegraphics[width=\columnwidth]{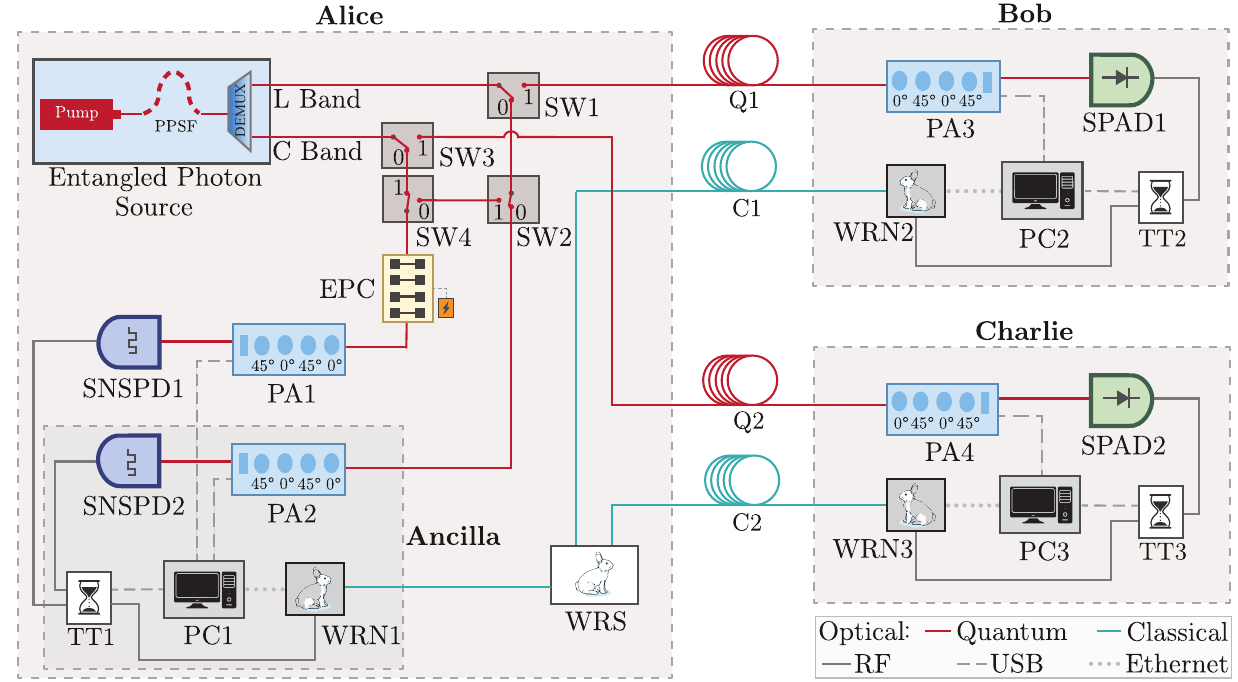}
    \caption{Schematic of the experimental setup. Details are provided in the text.
    EPC: electronic polarization controller; PA: polarization analyzer; PC: personal computer; PPSF: periodically poled silica fiber; SNSPD: superconducting nanowire single-photon detector; SPAD: single-photon avalanche diode; SW: switch; TT: time tagger; WRN: White Rabbit node; WRS: White Rabbit switch.}
    \label{fig:expSetup}
\end{figure}

\Cref{fig:expSetup} depicts the experimental setup.
Quantum and classical signals are transmitted via single-mode optical fibers in the dropped ceiling (Link Q and Link C), which connect Alice to Bob (Q$1$, C$1$; $86$~m; $1.4$~dB loss), and Alice to Charlie (Q$2$, C$2$; $60$~m; $2$~dB loss).
A four-port switching system---represented schematically by SW1--SW4 in \cref{fig:expSetup}, but implemented experimentally via manual fiber connections---manages signal routing across the network.

\begin{table}[!b]  
\caption{\label{tab:routing_configs} State preparation via optical routing through four optical switches (see Fig \ref{fig:expSetup}). Ports $0$ and $1$ denote the two output branches of each switch.}
\begin{ruledtabular}
\begin{tabular}{c|cccc}
\multicolumn{1}{c|}{State} & \multicolumn{4}{c}{Switch position} \\
$\rho_{ij}$ & SW 1 & SW 2 & SW 3 & SW 4 \\
\colrule
$\rho_{0A}$ & 0 & 0  & 0 & 1  \\
$\rho_{BA}/\rho_{BA'}$ & 1 & -- & 0 & 1  \\
$\rho_{BC}$ & 1 & -- & 1 & -- \\
$\rho_{AC}/\rho_{A'C}$ & 0 & 1  & 1 & 0  \\
\end{tabular}
\end{ruledtabular}
\end{table}
Polarization-entangled photon pairs are generated at Alice via continuous-wave spontaneous parametric down-conversion (SPDC) in a periodically poled silica fiber (PPSF).
The signal and idler are spectrally separated into the C and L telecom bands by a demultiplexer centered at $1567.3$~nm ($191.28$~THz) and distributed to their respective nodes.
Controlled birefringence is introduced to the Alice channel $\cE_A$ via an electronic polarization controller (EPC; OZ Optics EPC-$400$) driven by piezoelectric actuators.

At each node, incoming photons pass through polarization analyzers (PA$1$--PA$4$; Nucrypt PA-$1000$) before detection.
At Alice, superconducting nanowire single-photon detectors (SNSPD$1$, SNSPD$2$; Quantum Opus) provide $\geq$80\% detection efficiency with a $50$~ns dead time; at Bob and Charlie, free-running single-photon avalanche diodes (SPAD$1$, SPAD$2$; IDQuantique) operate at $25\%$ efficiency with a $10$~$\upmu$s dead time.
Detection events are logged by time taggers (TT$1$--TT$3$; Swabian TTU-$1024$) and synchronized via White Rabbit nodes (WRN$1$--WRN$3$; Orolia) linked to the White Rabbit Switch (WRS; Orolia) at Alice over Link C.
Detectors, PAs, and WRNs interface with the time taggers, local computers (PC$1$–PC$3$), and the WRS via RF, USB, and Ethernet connections, respectively, enabling centralized instrument control from PC$1$ at Alice.

For each configuration in \cref{tab:routing_configs}, $S=36$ joint polarization projections spanning the rectilinear $\{\ket{H},\ket{V}\}$, diagonal $\{\ket{D},\ket{A}\}$, and circular $\{\ket{R},\ket{L}\}$ bases are performed, with coincidences within a $1$~ns window accumulated over $20$~s.
As a baseline, the input state $\rho$---ideally the Bell state $\ket{\psi^{+}} =\frac{1}{\sqrt{2}}(\ket{HV}+\ket{VH})$---is measured locally at Alice's node to obtain the estimate $\hrho$.
The degree of birefringence is set by operating the EPC in full scrambling mode over a chosen duty cycle.
For example, to create $\cE_A^\star$ in \cref{eq:Aideal} with $p=0.2$, the EPC is turned on for 4~s and left off for 16~s of the total integration time.
\begin{figure}[!tb]\centering
\includegraphics[width=\columnwidth]{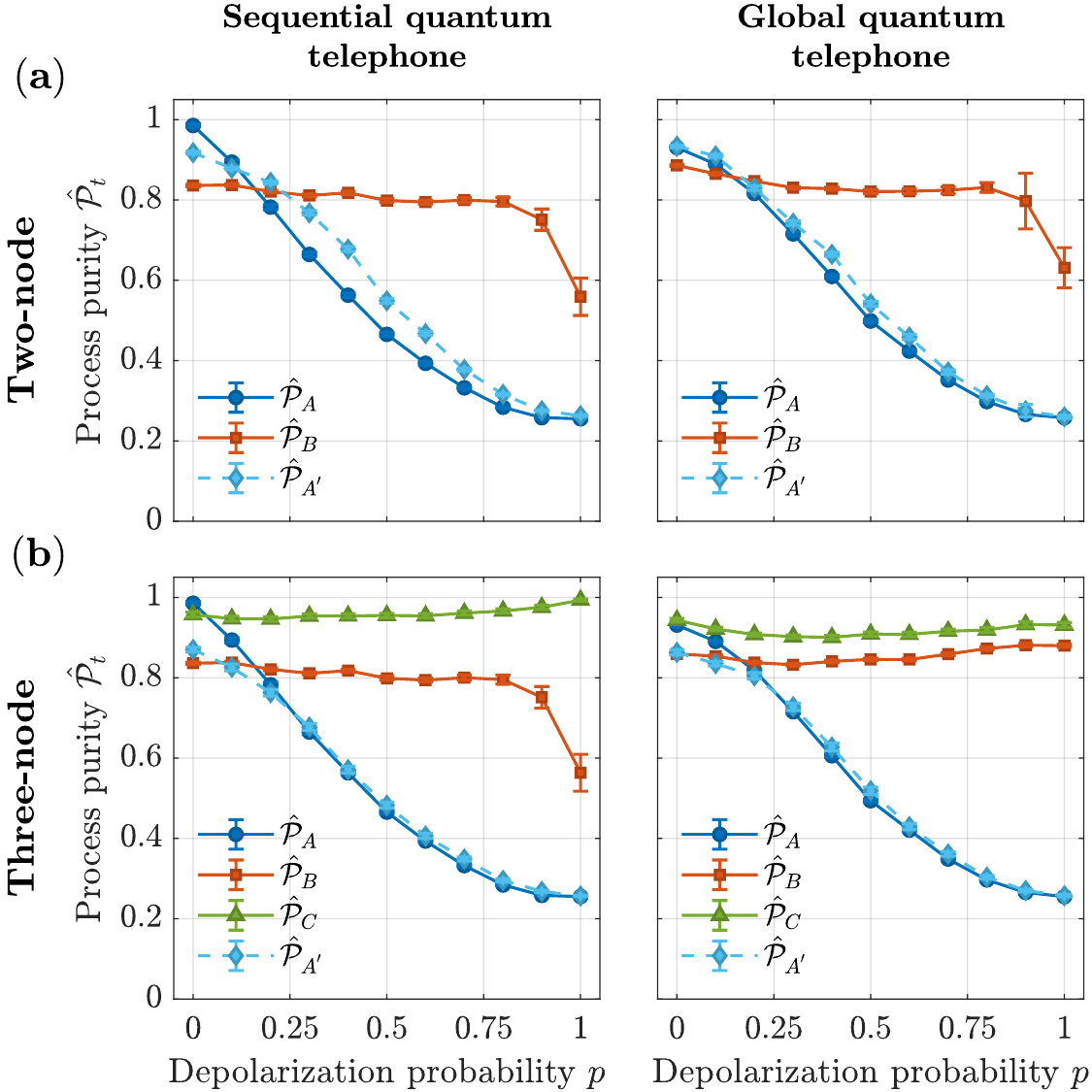}
    \caption{Process purities $\hcP_A$, $\hcP_B$, $\hcP_C$, and $\hcP_{A'}$ as a function of depolarization probability $p$ of the targeted ground truth Alice channel $\cE_A^\star$.
    (a)~Two-node configuration. (b) Three-node configuration.
    Left and right columns delineate between sequential inference via \cref{eq:sequential} and global inference via \cref{eq:global}.
    Channels $\hcE_A$ and $\hcE_{A'}$ should be compared to a depolarizing channel [\cref{eq:Aideal}], while   $\hcE_B$ and $\hcE_C$ should be compared to unitary channels [\cref{eq:BCideal}].} 
    \label{fig:process}
\end{figure}

Estimated purities for the two-node configuration are summarized in \cref{fig:process}(a).
As expected for \emph{bona fide} AAPT, inference of $\hcE_A$ closely matches the ground truth, with a purity $\hcP_A$ that follows theoretical expectations.
The bootstrapped estimate $\hcE_B$ returns a high flat purity $\hcP_B$ until $p>0.8$, after which point the decorrelating effects of Alice's channel begin to reduce the information gleaned from the probing of $\cE_B$, and the inferred purity drops.
The purity of $\hcE_{A'}$ follows closely behind that of $\hcE_{A'}$ and does not noticeably deviate from the ground truth even when the upstream $\hcE_B$ does. 
This follows from $\cE_A^\star$ being itself depolarizing, so that the ground truth and mean of the uniform Bayesian prior coalesce at $p=1$. 
Therefore what might otherwise appear to be a paradoxical persistence of information is more of a statistical coincidence: the depolarizing noise naturally aligns the degraded data with the uniform Bayesian prior, producing accurate estimated purity $\hcP_{A'}$ even when empirical information is lacking. 
Nonetheless, this result highlights by example the possibility for more accurate inference of bootstrapped channels than of upstream ones (i.e., $|\hcP_t-\cP_t^\star|<|\hcP_{t-1}-\cP_{t-1}^\star|$). 

Overall, both sequential and global approaches return similar results in the two-node experiment.
Yet this picture changes remarkably for three nodes.
Whereas \cref{fig:process}(b) confirms the same overall behavior for estimating $\cE_A$ and  $\cE_{A'}$ throughout---as well as the downward trend of $\hcP_B$ in the sequential approach---the global method is surprisingly able to infer $\cE_B$ with consistent purity for all $p\in[0,1]$.
In other words, the insertion of the downstream channel $\cE_C$ improves estimation of the upstream channel $\cE_B$, suggesting important informational value beyond the simple intuitive picture of bootstrapping in \cref{eq:sequential}.

\begin{figure}[!tb]\centering
\includegraphics[width=\columnwidth]{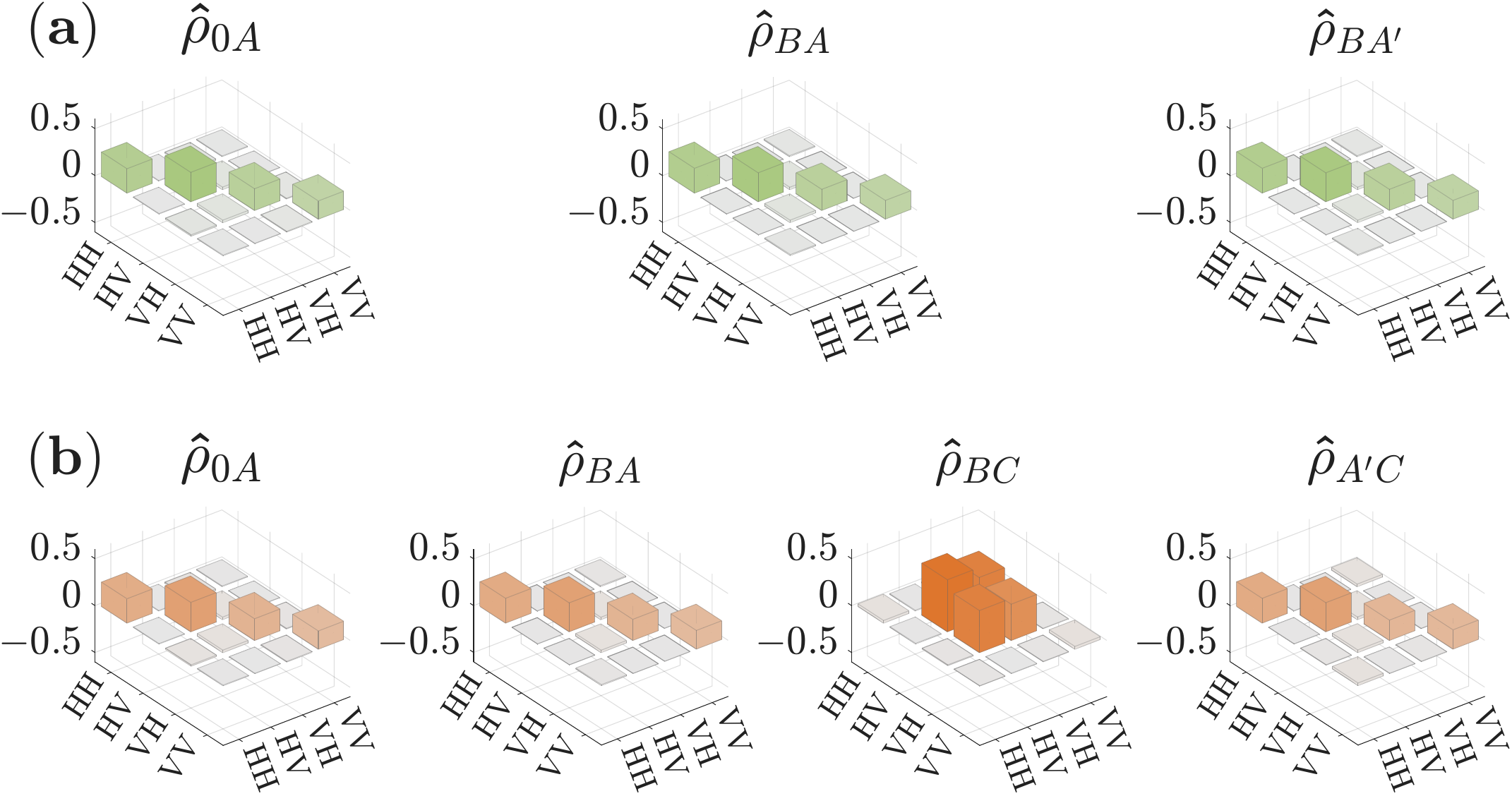}
    \caption{Real parts of the two-qubit density matrices $\hrho_{ij}$ estimated directly from telephone coincidence counts, for $p=1$ depolarizing probability on channel $\cE_{A(A')}^\star$. (a) Two-node telephone experiment. (b) Three-node telephone experiment.}
    \label{fig:state}
\end{figure} 
Insights into this paradoxical purity revival can be gained by reexamining the datasets from a QST perspective.
Although quantum telephone specifically focuses on estimating an unknown channel $\cE_t$, the same counts $\bN_t$ are sufficient on their own to estimate $\rho_{t-1,t}$ in \cref{eq:recursive} directly---a more straightforward task, for QST makes no attempt to extract the underlying processes $\cE_{t-1}\otimes\cE_t$. 
\Cref{fig:state} plots the estimated quantum states obtained from this procedure for the depolarizing extreme $p=1$. In the two-node case [\cref{fig:state}(a)], all three output states display the diagonal $\hrho\approx\frac{I_4}{4}$ associated with maximal mixedness, so although $\cE_B$ is \emph{in fact} unitary, the tomographic data offer no evidence in support thereof.

In contrast, the QST results for the three-node tests [\cref{fig:state}(b)] include one quantum state ($\hrho_{BC}$) that matches the maximally entangled state $\ket{\psi^+}$ with fidelity $\hcF_{BC}=0.960(4)$.
Since the only local channels that can preserve bipartite entanglement are unitary, this additional dataset implies that both $\cE_B$ and $\cE_C$ must be unitary; therefore, $\cE_B$ can now be inferred with high purity.
These states also explain why only \emph{global} quantum telephone fully unlocks this capability.
The sequential case finalizes its estimate $\hcE_B$ from only the count data corresponding to the upstream states $\hrho_{0A}$ and $\hrho_{BA}$, neither of which provides evidence of entanglement preservation.

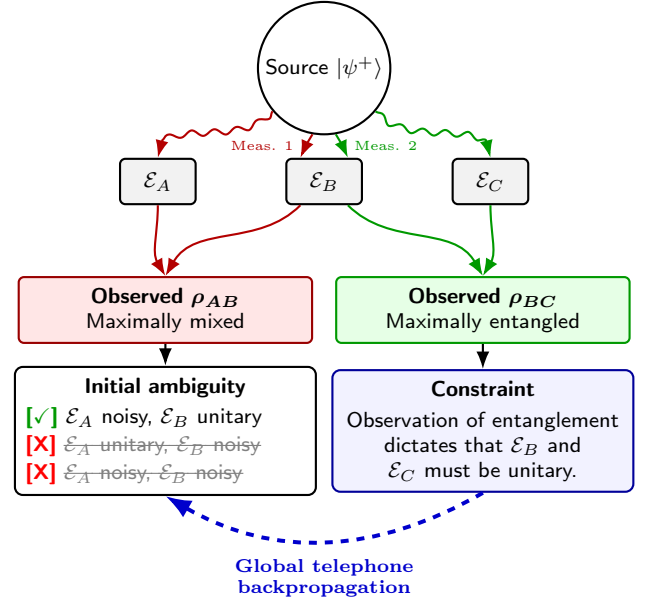
\begin{figure}[tb!] 
\centering
\begin{tikzpicture}[
    >=Latex,
    font=\sffamily\footnotesize,
    channel/.style={draw, thick, fill=gray!10, minimum width=1cm, minimum height=0.6cm, rounded corners=2pt, align=center},
    source/.style={circle, draw, thick, inner sep=2pt, align=center},
    output/.style={draw, thick, rounded corners=2pt, align=center, text width=3.6cm, inner sep=4pt},
    hypobox/.style={draw, thick, fill=white, rounded corners=3pt, align=left, text width=3.7cm, inner sep=4pt},
    constbox/.style={draw, thick, fill=blue!5, rounded corners=3pt, align=center, text width=3.7cm, inner sep=4pt},
    photon1/.style={decorate, decoration={snake, amplitude=0.4mm, segment length=3mm, post length=2mm}, ->, thick, draw=red!70!black},
    photon2/.style={decorate, decoration={snake, amplitude=0.4mm, segment length=3mm, post length=2mm}, ->, thick, draw=green!60!black}
]

    \node[source] (src) at (0, 0) {Source $\ket{\psi^+}$};
    
    \node[channel] (E1) at (-2.2, -1.5) {$\mathcal{E}_A$};
    \node[channel] (E2) at (0, -1.5) {$\mathcal{E}_B$};
    \node[channel] (E3) at (2.2, -1.5) {$\mathcal{E}_C$};

    \draw[photon1] (src.-140) to[out=-135, in=90] (E1.north);
    \draw[photon1] (src.-100) to[out=-90, in=120] node[pos=0.45, left=1pt, font=\tiny, text=red!70!black] {Meas. 1} (E2.135);
    
    \draw[photon2] (src.-80) to[out=-90, in=60] node[pos=0.45, right=1pt, font=\tiny, text=green!60!black] {Meas. 2} (E2.45);
    \draw[photon2] (src.-40) to[out=-45, in=90] (E3.north);

    \node[output, fill=red!10, draw=red!60!black] (obs1) at (-2.1, -3.2) {\textbf{Observed $\bm{\rho_{AB}}$}\\Maximally mixed};
    \node[output, fill=green!10, draw=green!60!black] (obs2) at (2.1, -3.2) {\textbf{Observed $\bm{\rho_{BC}}$}\\Maximally entangled};

    \draw[->, thick, draw=red!70!black] (E1.south) to[out=-90, in=120] (obs1.north);
    \draw[->, thick, draw=red!70!black] (E2.225) to[out=-135, in=60] (obs1.north);
    
    \draw[->, thick, draw=green!60!black] (E2.315) to[out=-45, in=120] (obs2.north);
    \draw[->, thick, draw=green!60!black] (E3.south) to[out=-90, in=60] (obs2.north);

    \node[hypobox, draw=black] (hypo) at (-2.1, -4.8) {
        {\centering\textbf{Initial ambiguity}\\[2pt]}
        \textcolor{green!60!black}{\textbf{[\checkmark]}} $\mathcal{E}_A$ noisy, $\mathcal{E}_B$ unitary\\[1pt]
        \textcolor{red}{\textbf{[X]}} \textcolor{gray}{\sout{$\mathcal{E}_A$ unitary, $\mathcal{E}_B$ noisy}}\\[1pt]
        \textcolor{red}{\textbf{[X]}} \textcolor{gray}{\sout{$\mathcal{E}_A$ noisy, $\mathcal{E}_B$ noisy}}
    };
    \node[constbox, draw=blue!60!black] (const) at (2.1, -4.8) {
        \textbf{Constraint}\\[2pt]
        Observation of entanglement dictates that \textbf{$\mathcal{E}_B$} and \textbf{$\mathcal{E}_C$} must be unitary.
    };

    \draw[->, thick] (obs1.south) -- (hypo.north);
    \draw[->, thick] (obs2.south) -- (const.north);

    \draw[->, dashed, line width=1.5pt, color=blue!80!black] 
        (const.south) to[out=-150, in=-30] 
        node[midway, below=3pt, font=\scriptsize\bfseries, color=blue!80!black, align=center] {Global telephone\\backpropagation} 
        (hypo.south);
\end{tikzpicture}
\caption{Resolving channel ambiguities through global quantum telephone.
Observing a maximally mixed output state $\rho_{AB}=(\cE_{A}\otimes\cE_B)(\ket{\psi^+}\bra{\psi^+})$ leaves ambiguity as to the quality of $\cE_A$ and $\cE_B$.
However, seeing a maximally entangled $\rho_{BC}=(\cE_{B}\otimes\cE_C)(\ket{\psi^+}\bra{\psi^+})$ requires that both $\cE_B$ and $\cE_C$ are unitary channels, information that can then be backpropagated to reveal only $\cE_A$ as noisy.}
\label{fig:sudoku}
\end{figure}

This intuitive explanation is summarized in \cref{fig:sudoku}, where we envision the most extreme form of this phenomenon: $\rho_{AB}=(\cE_{A}\otimes\cE_B)(\ket{\psi^+}\bra{\psi^+})$ is maximally mixed and $\rho_{BC}=(\cE_{B}\otimes\cE_C)(\ket{\psi^+}\bra{\psi^+})$ is maximally entangled.
The first observation could be obtained by depolarization from either or both $\cE_A$ and $\cE_B$, leaving an initial ambiguity in the quality of each channel. But the second observation is only possible when both $\cE_B$ and $\cE_C$ are unitary, information which resolves the initial ambiguity and confirms that only $\cE_A$ is noisy.

\section{Discussion}
Because quantum telephone is based on estimating $T$ unknown one-qudit channels $\{\cE_t\}$ from $T$ two-qudit states $\{\rho_{t-1,t}\}$, it does not reduce the total number of measurements compared to standard AAPT.
Accordingly, whenever it is possible to probe each unknown quantum channel in isolation, there is no real advantage in implementing a telephone-type experiment over conventional process tomography.
However, as quantum networks scale, it is expected that many nodes will lack the resources for independent channel characterization.
In these contexts, quantum telephone principles might prove the only viable path to extract information about unknown quantum channels, particularly when pairwise entanglement has already been established.

The features observed in this experiment suggest further tomographic variations in the spirit of quantum telephone, such as the intentional insertion of pre-characterized channels to aid estimation of unknown ones, or the inference of cascaded quantum channels---e.g., $\rho_t=(\mathds{1}\otimes\cE_t\circ\cdots\circ\cE_0)(\rho)$---instead of the parallel ones in \cref{eq:recursive}.
In any of these directions, we expect that the global flavor of quantum telephone [\cref{eq:global}] will continue to prove essential in enforcing all relevant physical and logical constraints, thereby removing ambiguities from single observations in the manner as highlighted in \cref{fig:sudoku}.
For such problems, the ability of likelihood-based methods to handle arbitrary probabilistic frameworks makes maximum likelihood estimation (MLE)~\cite{Hradil1997,James2001,Lvovsky2004} and Bayesian inference~\cite{Blume-Kohout2010,Lukens2020b, Nguyen2025} leading candidates from the algorithmic side, with the selection depending on the tradeoff between computational speed (MLE preferred) and theoretical optimality (Bayesian inference preferred).
Our demonstration of simultaneous Bayesian estimation of four unknown quantum channels---to our knowledge, the first time AAPT has been chained to jointly infer multiple distinct quantum processes from a single global dataset, Bayesian or otherwise---shows the feasibility of global quantum telephone even with computationally demanding MCMC sampling.

Nonetheless, as more channels are considered, the $\mathcal{O}(Td^4)$ growth in the parameter space will eventually limit computational feasibility.
Future work must explore hybrid or localized inference models that can incorporate downstream boundary constraints without requiring full global computation.
Ultimately, the quantum telephone framework reveals that cascaded quantum channels contain hidden tomographic information that, when properly extracted, establishes a practical path for characterizing complex network architectures.

\begin{acknowledgments}
This work was performed in part at Oak Ridge National Laboratory, operated by UT-Battelle for the U.S. Department of Energy (DOE) under contract DE-AC05-00OR22725. Funding was provided by DOE (ERKJ432).
\end{acknowledgments}

\textit{Data availability.---}The data that support the findings of this letter are openly available~\cite{arefur_rahman_2026_21358142}.

\appendix
\section{Fidelity-based analysis}
\label{sec:fidelityAnalysis}
In the main text we focus on Choi matrix purity $\cP=\Tr\Phi^2$ as the primary metric of analysis because of its imperviousness to local rotations: $\cP=1\Longleftrightarrow\cE(\sigma)=U\sigma U^\dagger$ for any unitary $U$.
Since such local rotations are generally unavoidable in deployed fiber due to random birefringence---and compensated in practice by fixed polarization controllers---this choice allows us in quantum telephone to concentrate on more fundamental questions related to noise and unitarity.
Nevertheless, it is possible to leverage channel fidelity in a similar fashion, subject to gauge nuances discussed below.

The fidelity between a general quantum process $\Phi$ and a target $\Phi^\star$ is given by
\begin{equation}
\label{eq:fidelityChoi}
\cF = \left(\Tr\sqrt{\sqrt{\Phi^\star}\Phi\sqrt{\Phi^\star}}\right)^2.
\end{equation}
If $\Phi^\star$ is unitary so that $\cE^\star(\sigma)=U^\star\sigma(U^\star)^\dagger$, \cref{eq:fidelityChoi} can be written in terms of the Kraus operators $\{A_k\}_{k\in[R]}$ associated with $\Phi$ as~\cite{Johnston2011}
\begin{equation}
\label{eq:fidelityKraus}
\cF = \frac{1}{d^2}\sum_{k=1}^R \left|\Tr A_k^\dagger U^\star \right|^2.
\end{equation}
Now, if $\Phi$ describes a unitary process as well, we can choose a rank-1 Kraus representation such that $A_k=U\delta_{k1}$, and \cref{eq:fidelityKraus} simplifies to $\cF =\frac{1}{d^2}|\Tr U^\dagger U^\star |^2$---the Hilbert--Schmidt inner product common in linear-optical quantum computing~\cite{Uskov2009, Clements2016, Lukens2017}.

Yet when $U$ deviates strongly from $U^\star$, the fidelity can be extremely low (even zero), while in most practical networking contexts such a unitary channel would nevertheless be ideal.
In order to let fidelity reflect such perfection, it is common to introduce a fixed unitary rotation $V$ ($A_k\rightarrow VA_k$) and define the fidelity as the maximum over elements of the set of $d$-dimensional unitaries $\mathbb{U}(d)$:
\begin{equation}
\label{eq:fidelityVrot}
\cF =\max_{V\in\mathbb{U}(d)} \frac{1}{d^2}\sum_{k=1}^R \left|\Tr A_k^\dagger V^\dagger U^\star \right|^2,
\end{equation}
a procedure analogous to the fully entangled fraction defined in quantum state contexts~\cite{Grondalski2002}.
Consequently, we see no hesitation in calling \cref{eq:fidelityVrot} ``the'' process fidelity for a unitary target $U^\star$ in contexts like quantum networking where any fixed $V$ is expected and compensated.

However, additional nuances appear in Bayesian process tomography, which returns a set of possible processes $\{\Phi(\bx^i)\}_{i\in[N]}$.
The uncompensated Bayesian mean fidelity is the average
\begin{equation}
\hcF = \frac{1}{Nd^2}\sum_{i=1}^N\sum_{k=1}^R \left|\Tr A_k^\dagger(\bx^i)U^\star \right|^2,
\end{equation}
and so the principle behind \cref{eq:fidelityVrot} can be applied in two different ways, depending on whether the rotation performed is fixed or varies sample by sample:
\begin{widetext}
\begin{equation}
\label{eq:twoWays}
\hcF = 
\begin{dcases}
\max_{V\in\mathbb{U}(d)} \frac{1}{Nd^2}\sum_{i=1}^N\sum_{k=1}^R \left|\Tr A_k^\dagger(\bx^i)V^\dagger U^\star \right|^2 & ;\quad \text{fixed unitary reference} \\
\frac{1}{Nd^2}\sum_{i=1}^N 
\max_{V(\bx^i)\in\mathbb{U}(d)}\sum_{k=1}^R \left|\Tr A_k^\dagger(\bx^i)V^\dagger(\bx^i) U^\star \right|^2  & ;\quad \text{variable unitary reference}
\end{dcases}.
\end{equation}
\end{widetext}
The fidelity computed by the first definition is strictly less than or equal to that of the second.
Indeed, in the extreme case of unitary processes drawn from the Haar distribution, $\hcF=\frac{1}{d^2}$ for the first definition while $\hcF=1$ in the second.
Conceptually, the fixed definition compares all feasible samples to a single ideal unitary process, while the variable definition compares each sample to its own bespoke unitary process.

So which of the two definitions in \cref{eq:twoWays} should be embraced by quantum telephone?
The more conservative fixed rotation is certainly easy to justify.
Yet we would argue that the second definition is reasonable as well, and in fact even \emph{more} appropriate for channel estimation in deployed quantum networks.
To this end, it is important to emphasize that the posterior distribution in Bayesian inference summarizes knowledge over mutually exclusive possibilities: each sample $\bx^i$ corresponds to a complete quantum process whose physical realization precludes that of a different sample $\bx^{j\neq i}$.
Thus a posterior distribution equal to the Haar distribution does not imply the ground truth is a mixture; rather, it implies that the ground truth is some fixed---though as yet unknown---unitary matrix.
This conclusion carries significant experimental ramifications, for it means that an empirically chosen local rotation can convert the channel to unit fidelity.

In light of these considerations, the variable-reference fidelity definition in \cref{eq:twoWays} is arguably more informative than the first definition, as it quantifies the average attainable fidelity under quantum process uncertainties. Nevertheless, to avoid the potential for confusion in explaining these differences, we focus on the purity in the main text, which is automatically invariant to all local rotations.
\begin{figure*}[!tb]
\centering
\includegraphics[width=\textwidth]{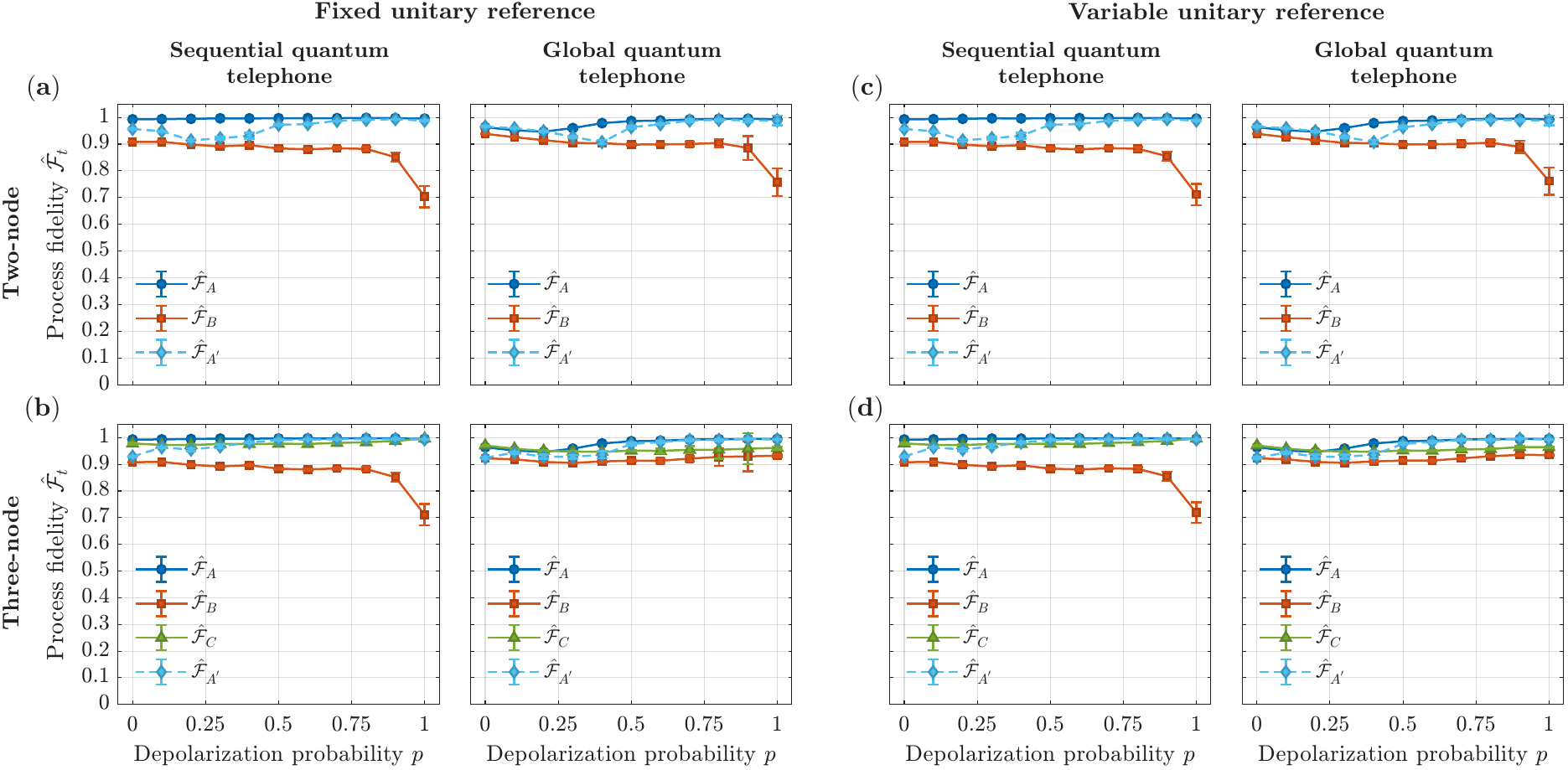}
    \caption{Experimentally obtained process fidelities $\hcF_A$, $\hcF_B$, $\hcF_C$, and $\hcF_{A'}$ as a function of depolarization probability of the ground truth Alice channel $\cE_A^\star$.
    (a,b)~Fixed unitary correction for all Bayesian samples.
    (c,d)~Bespoke unitary correction for each Bayesian sample.
    (a,c)~Two-node configuration.
    (b,d) Three-node configuration.
    Subsequent columns delineate between sequential inference and global inference.}
    \label{fig:exp_process_fidelity}
\end{figure*}

For completeness, \cref{fig:exp_process_fidelity} plots the Bayesian-estimated fidelities of the experimental two- and three-node results, using both rotation methods.
For a generic target whose Choi matrix $\Phi^\star$ is mixed---i.e., the cases $\cE_{A(A')}^\star$---the two rotation methods can be summarized as
\begin{widetext}
\begin{equation}
\label{eq:twoWaysMixed}
\hcF = 
\begin{dcases}
\max_{V\in\mathbb{U}(d)} \frac{1}{N}\sum_{i=1}^N \left\{ \Tr \sqrt{\sqrt{\Phi^\star}[I_d\otimes V]\Phi(\bx^i)[I_d\otimes V^\dagger]\sqrt{\Phi^\star}}\right\}^2 & ;\quad \text{fixed unitary reference} \\
\frac{1}{N}\sum_{i=1}^N 
\max_{V(\bx^i)\in\mathbb{U}(d)} \left\{ \Tr \sqrt{\sqrt{\Phi^\star}[I_d\otimes V(\bx^i)]\Phi(\bx^i)[I_d\otimes V^\dagger(\bx^i)]\sqrt{\Phi^\star}}\right\}^2  & ;\quad \text{variable unitary reference}
\end{dcases},
\end{equation}
\end{widetext}
which reduces to \cref{eq:twoWays} when $\Phi^\star$ is pure.
Overall, the fidelity results in \cref{fig:exp_process_fidelity}(a,b) confirm the findings of the purities plotted in \cref{fig:process}(a,b): estimation of $\cE_A$, $\cE_C$, and $\cE_{A'}$ shows good agreement with theory for all depolarization probabilities $p$, while the fidelity $\hcF_B$ drops as $p\rightarrow 1$ for all cases \emph{except} global inference of the three-node test, due to the availability and use of downstream high-purity results.

Interestingly, the variable-reference fidelities in \cref{fig:exp_process_fidelity}(c,d) do not show significant differences from the fixed-reference ones [\cref{fig:exp_process_fidelity}(a,b)], indicating that the posterior distribution is sufficiently clustered such that the questions surrounding  Haar-random averaging above do not apply---a situation that will change in the simulations described below.

\section{Simulations}
\label{sec:simulations}
In this section, we conduct numerical simulations of the quantum telephone experiments described in the main text, employing both sequential and global strategies on two-node and three-node network configurations.
Coincidence data are generated randomly  from a Poisson distribution, where the rate parameters are defined as the product of the mean integrated coincidence rate $K$ and the measurement probabilities $\wp_{ts}$ introduced in \cref{eq:seqProb}, using the ideal channels in \cref{eq:BCideal,eq:Aideal} with $U=I_2$ for concreteness.
Specifically, we employ mean count values $K$ of $6000$, $5500$, and $5000$ corresponding to the output states $\rho_{0A}$, $\{\rho_{BA}, \rho_{BA^{'}}, \rho_{A^{'}C}\}$, and $\rho_{BC}$ as specified in \cref{eq5:qt_maps_2n,eq6:qt_maps_3n}, respectively, designed to closely match the mean counts measured experimentally.
We then perform inference on these simulated results according to the identical Markov chain Monte Carlo (MCMC) workflow applied to the experimental results.

\begin{figure*}[!tb]\centering
\includegraphics[width=\textwidth]{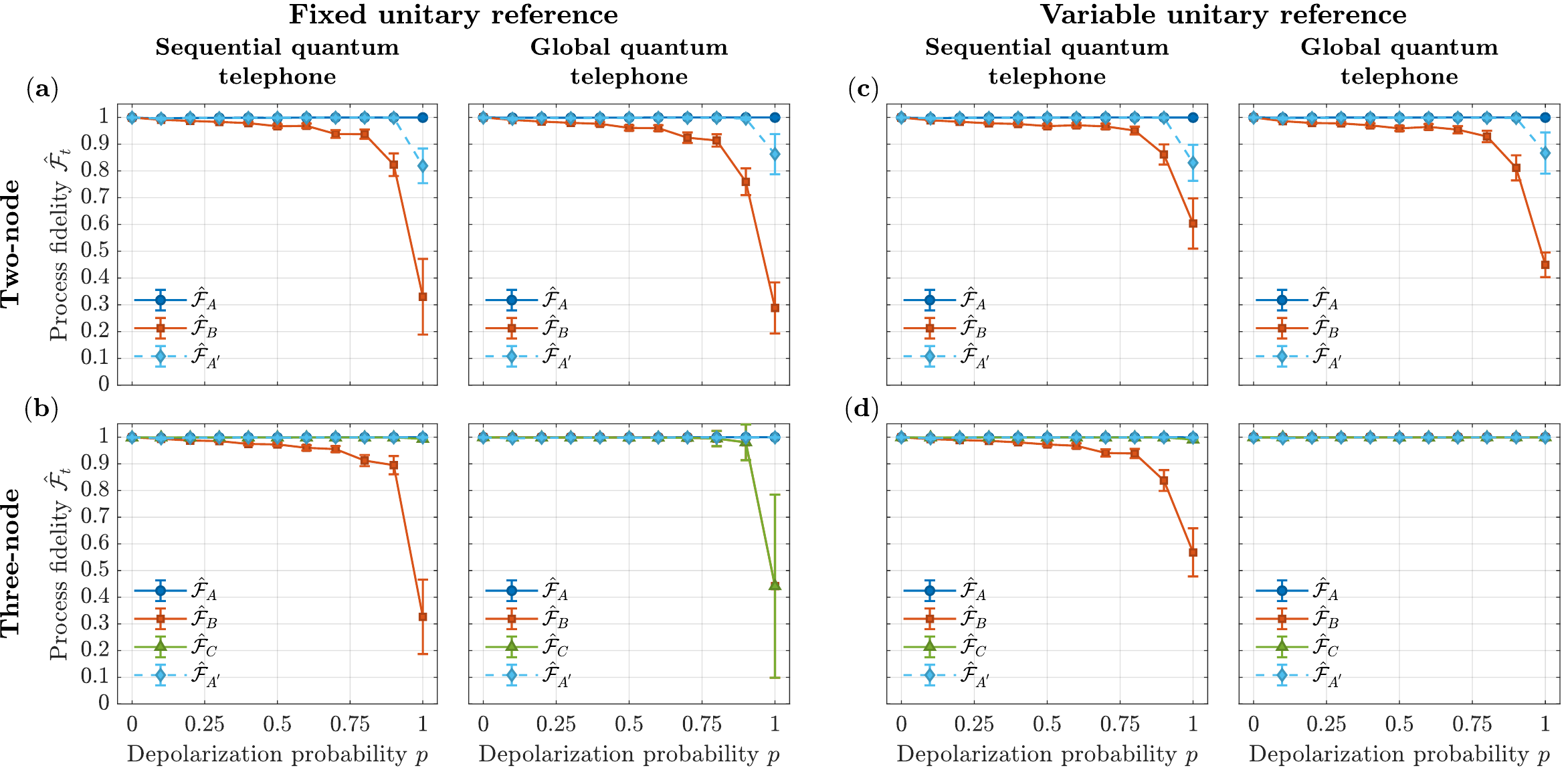}
    \caption{Numerical simulation of process fidelities $\hcF_A$, $\hcF_B$, $\hcF_C$, and $\hcF_{A'}$ as a function of depolarization probability of the ground truth Alice channel $\cE_A^\star$. (a,b)~Fixed unitary correction for all Bayesian samples. (c,d)~Bespoke unitary correction for each Bayesian sample. (a,c)~Two-node configuration. (b,d) Three-node configuration. Subsequent columns delineate between sequential and global inference.}
    \label{fig:sim_process_fidelity}
\end{figure*}

\Cref{fig:sim_process_fidelity}(a,b) plots the process fidelities obtained in simulation using the first, fixed cases of Eqs.~(\ref{eq:twoWays},\ref{eq:twoWaysMixed}) for unitary correction.
The trends follow those of the experimental results in \cref{fig:exp_process_fidelity}(a,b), except for higher fidelities at low $p$ and more pronounced drops at $p=1$ for $\hcF_{B}$,  $\hcF_{C}$, and $\hcF_{A'}$.
While obtaining higher fidelities for small $p$ is unsurprising for ideal simulations, the lower simulated fidelities at $p=1$ demand further attention.
Applying \emph{variable} unitary correction, we obtain the results in \cref{fig:sim_process_fidelity}(c,d), where the $\hcF_C$ dip is eliminated entirely and the revival of $\hcF_B=1$ at $p=1$ is obtained for global quantum telephone, in line with the experimental results.

These simulation results therefore prompt two main questions:
(i) How can the presence of the sharp drops in $\hcF_{B(C)}$ in \cref{fig:sim_process_fidelity}(b) and their removal in \cref{fig:sim_process_fidelity}(d) be reconciled with the concept of channel information recovery described in the main text and summarized in Fig.~5?
(ii) Why was this drop not observed experimentally in the right column of \cref{fig:exp_process_fidelity}(b)?

The first question is related to the discussion in the previous section---namely, the difference between knowing a channel is unitary and knowing which specific unitary describes the channel.
The former knowledge is sufficient for high fidelity with respect to a variable unitary reference [measured in \cref{fig:sim_process_fidelity}(d)], while the latter is required for high fidelity with respect to a fixed unitary reference [measured \cref{fig:exp_process_fidelity}(b)].
Mathematically, the joint observation of low fixed fidelity but high variable fidelity can be understood from the well-known ``transpose trick'' for maximally entangled states~\cite{Wilde2017}.
Suppose that a maximally entangled state $\ket{\psi}$, defined by some rotation $Y$ on the Bell state $\ket{\phi^+}=\frac{1}{\sqrt{d}}\sum_{n=1}^d\ket{nn}$ as
\begin{equation}
\ket{\psi} = (I_d\otimes Y)\ket{\phi^+},
\end{equation}
is sent through two unitary channels $U$ and $V$ such that it appears unchanged at the output:
\begin{equation}
(U\otimes V)\ket{\psi} = \ket{\psi}.
\end{equation}
This outcome is possible for any pair of unitaries $(U,V)$ related by $V=YU^*Y^\dagger$.
Consequently, while observation of a maximally entangled output---similar to $\hrho_{BC}$ in \cref{fig:state}---implies that both $\cE_B$ and $\cE_C$ are unitary, it specifies their values only up to a local gauge $U$.
Therefore the Choi states $\Phi(\bx)$ sampled by MCMC fill the surface of a high-dimensional Bloch sphere, averaging toward $\hcF=\frac{1}{d^2}$ with respect to any fixed unitary reference but achieving $\hcF=1$ with a sample-dependent unitary reference.

That leaves us with the second question: why do we not see such a drop in the fixed-reference fidelity in the experimental results of \cref{fig:exp_process_fidelity}(b)?
Our leading hypothesis attributes this outcome to the  experimental depolarization not reaching true $p=1$ in the full-scrambling case.
Numerically, we find that a depolarizing channel $\cE_A^\star$ with $p\approx0.92$ gives slightly higher fidelity with respect to $\hcE_A$ in the full-scrambling cases, suggesting that indeed we are not fully reaching the nominal target $p=1$.
Regardless of these questions, the use of fidelity optimized over a variable unitary reference [Figs.~\ref{fig:exp_process_fidelity}(d) and \ref{fig:sim_process_fidelity}(d)] yields broad agreement between experiment and simulation, confirming the basic conceptual understanding of quantum telephone's use of tomographic information in estimating quantum channels.

\bibliography{references}

\end{document}